\documentclass[preprint]{aastex}

\newcommand{\fex}{Fe~{\sc x}}
\newcommand{\fexi}{Fe~{\sc xi}}
\newcommand{\fexiii}{Fe~{\sc xiii}}
\newcommand{\fexiv}{Fe~{\sc xiv}}

\shorttitle{Thermal structure of coronal loops}
\shortauthors{S. K. Prasad, J. Singh and Ichimoto}

\begin{document}
\title{Thermal structure of coronal loops as seen with Norikura coronagraph}             
\author{S. Krishna Prasad and Jagdev Singh} 
\affil{Indian Institute of Astrophysics, II Block, Koramangala, Bangalore 560 034, India.}                                  
\email{krishna@iiap.res.in}
\author{K. Ichimoto} 
\affil{Kwasan and Hida Observatories, Kyoto University, Yamashina-ku, Kyoto 607-8417, Japan}                                  

\begin{abstract}
The thermal structure of a coronal loop, both along and across the loop, is vital in determining the exact plasma heating mechanism. High resolution spectroscopic observations of the off-limb corona were made using 25~cm Norikura coronagraph, located at Norikura, Japan. Observations on a number of days, were made simultaneously in four forbidden iron emission lines namely, [\fexi] 7892~\AA\ line, [\fexiii] 10747~\AA~ \& 10798~\AA\ lines and [\fexiv] 5303~\AA\ line and on some days made only in [\fexi] 7892~\AA\ and [\fex] 6374~\AA\ lines. Using the temperature sensitive emission line ratios [\fexiv] 5303~\AA/[\fexiii] 10747~\AA, and [\fexi] 7892~\AA/[\fex] 6374~\AA\ we compute the electron temperatures along 18 different loop structures observed on different days. We find a significant negative temperature gradient in all the structures observed in \fexiv\ and \fexiii\ and a positive temperature gradient in the structures observed in \fexi\ and \fex. Combining these results with the previous investigations by Singh and his collaborators, we infer that the loop tops, in general, appear hotter when observed in colder lines and colder when observed in relatively hotter lines as compared to their coronal foot points. We suggest that this contrasting trend observed in the temperature variation along the loop structures can be explained by a gradual interaction of different temperature plasma. The exact mechanism responsible for this interaction needs further quest and has potential to constrain loop heating models.
\end{abstract}
\keywords{Sun: activity---Sun: corona---methods: data analysis---techniques: spectroscopic}

\section{Introduction}
Plasma in the solar atmosphere is mostly confined to magnetic structures called coronal loops which are regarded as the basic building blocks of the solar corona. Any improvement in the knowledge on these loops, like the exact thermal structure of a coronal loop and other such properties, can therefore help us in solving several long-lasting issues related to corona (for \textit{e.g.} coronal heating). One of the earliest attempts to theoretically model the coronal loops, was by \citet{1978ApJ...220..643R}, which is widely known as RTV model. They derived scaling laws between temperature, pressure and length of the loop and demonstrated that in a uniformly heated stable hydrostatic loop the temperature maximum must be located near its apex. \citet{1981ApJ...243..288S} generalized this model to long loops by including the variation of pressure and heat deposition along the loop. \citet{1996PASJ...48..535K} derived temperature distribution in 16 steady loops observed with \textit{Yohkoh} in soft X-rays and found that the temperature is highest around the loop top and decreases towards the foot points consistent with these models. Similar studies on cooler EUV loops, however, showed a little or no temperature variation along the loop which is not in agreement with the static loop models \citep{1999ApJ...517L.155L, 2000ApJ...541.1059A}. Different theories were proposed to explain this isothermal nature. \citet{2000ApJ...528L..45R} suggested that a superposition of several unresolved thin strands at different temperatures can produce a flatter temperature profile in the observed loop. \citet{2001ApJ...550.1036A} proposed that for long EUV loops most of the heating is concentrated at the foot points resulting in the near isothermal loop structure. \citet{2002ApJ...567L..89W} showed that inclusion of flows in the coronal loop models (non-static loops) can give similar results. Non-uniformity in the loop cross section, with a significant decrease near the foot points can also make the temperature profile more isothermal than that in the case of constant cross section \citep{2004ApJ...611..537L}. Some other studies \citep{2003A&A...406.1089D, 2006A&A...449.1177R} disagree on the observed isothermal nature. Later studies mostly focused on the cross-field thermal structure of the loops. Here, we investigate the thermal profile along coronal structures using spectroscopic data observed with the 25~cm Norikura coronagraph. We give the observational details and data analysis methods in the following sections and discuss the results in the final section.

\section{Observations}
In view of the complex nature of intensity ratios observed \citep{2004ApJ...617L..81S}, further observations were planned and obtained by Jagdev Singh using 25~cm coronagraph for this study. The observations were carried out simultaneously in four forbidden iron emission lines namely, [\fexi] 7892~\AA\ line, [\fexiii] 10747~\AA~ \& 10798~\AA\ (IR) lines, and [\fexiv] 5303~\AA\ line, using three different CCD cameras (spectra in both the IR lines were imaged on a single CCD). In addition to these, some observations were made simultaneously in only two lines, [\fexi] 7892~\AA\ and [\fex] 6374~\AA. The instrumental setup is similar to that described in  \citet{1999PASJ...51..269S, 2003ApJ...585..516S}.
\begin{table}
\begin{center}
\caption{Details of observations}
\label{tbl1}
 \begin{tabular}{ccc}
 \tableline \tableline
  Date  & Target & Observation \\
        & region & time (JST\tablenotemark{a}) \\
 \tableline
  2005 Sep 21 & West limb & 09:28 -- 09:50 \\
  2005 Oct 3  & East limb & 08:28 -- 08:47 \\
  2005 Oct 6  & East limb & 06:05 -- 06:41 \\
  2007 Oct 6\tablenotemark{b} & West limb & 11:57 -- 12:29 \\
 \tableline
 \end{tabular}
\tablenotetext{a}{Japanese Standard Time}
\tablenotetext{b}{Observed only in two lines \fexi\ 7892~\AA\ and \fex\ 6374~\AA.}
\end{center}
\end{table}
Four different raster scans, with relatively better signal-to-noise, were chosen for this study. Observation time and the target region for these four scans are listed in Table~\ref{tbl1}. First three scans were observed simultaneously in four lines and the last scan was observed only in two lines. We refer to these two sets as set I and set II respectively, for the rest of the paper. The slit length was around 500\arcsec\ and the width was 5\arcsec\ for the scans in set I and 3\arcsec\ for those in set II. Each scan covers a portion of the equatorial off-limb corona of about 200\arcsec\ $\times$ 500\arcsec\ size. Exposure times were varied from day to day and also between different lines from 25~s to 50~s depending on the target of observation to get good signal. This translates to a duration of 20~min to 35~min to complete a single scan (see Table~\ref{tbl1}). Disk spectra were obtained immediately after or before the scan, by keeping the slit at the center of the solar disk. Corresponding dark and flat spectra were taken after each set of observations. Reference slit images (wire spectra) were also taken with the same setup by keeping three wires across the slit at a fixed known distance to correlate the spectral images obtained in different lines. 

\section{Data Reduction and Analysis}
The data were prepared following the standard reduction procedure for the Norikura datasets. Each spectral image had been corrected for the dark current and pixel to pixel variations using the corresponding dark and flat images. Scattered light component in the spectra was subtracted using the disk spectra. One of the scans in set I (taken on 2005 October 6) was binned on two pixels in spectral dimension to improve the signal-to-noise. The dispersion values at each CCD was computed by comparing the corresponding disk spectra with the standard solar spectra. The dispersion values for the scans in set I are 40.3~m\AA~pixel$^{-1}$, 74.2~m\AA~pixel$^{-1}$, and 19.3~m\AA~pixel$^{-1}$, respectively for \fexi, \fexiii, and \fexiv\ lines (twice the values for the binned case) and the values for those in set II are 23.6~m\AA~pixel$^{-1}$ and 26.6~m\AA~pixel$^{-1}$ respectively for \fexi\ and \fex\ lines. All the spectra were fitted with a single Gaussian profile and the parameters such as amplitude (peak counts), peak position, and line width, were derived. Only those spectra, with a signal-to-noise 5 or greater, are considered. Gaussian amplitudes were converted to absolute intensities using the disk spectra and standard solar flux values at respective wavelengths. The total area under the fitted Gaussian curve is taken as the total intensity in the line profile which is used in this analysis. Peak positions were converted to Doppler velocities taking the average position in the scan as reference. Line widths were converted to FWHM after correcting for the instrumental broadening as explained in \citet{2013SoPh..282..427P}. A 2-D monochromatic image of the observed region was then constructed in each of these parameters. Spatial scale of these images in each spectral line is not same due to different focal lengths of the optics used to focus the spectra. This is also affected by the differences in pixel sizes. 
\begin{figure}
\centering
\includegraphics[scale=1.0]{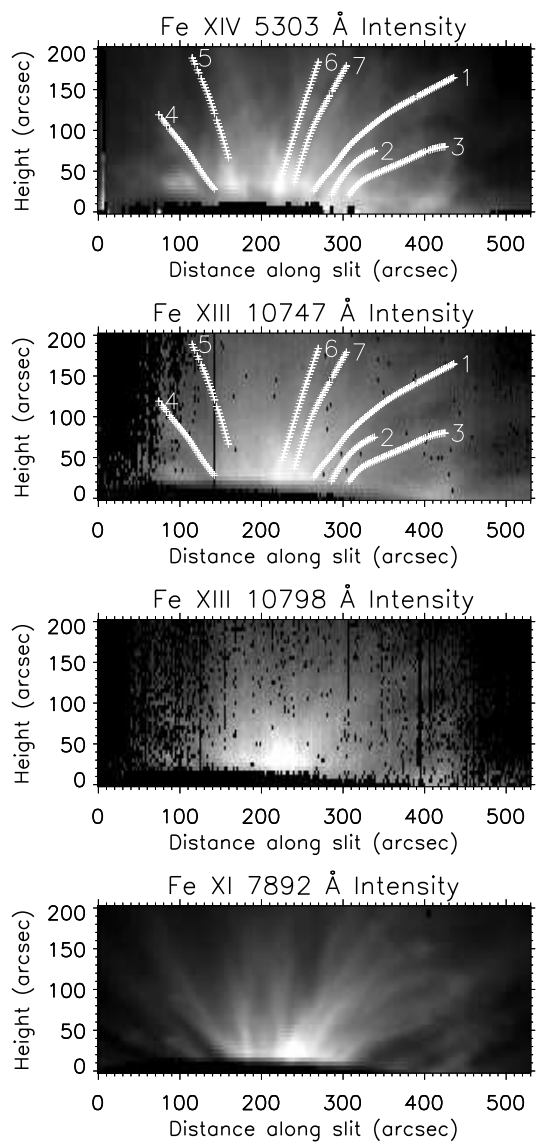}
\caption{Intensity images in \fexiv\ 5303~\AA, \fexiii\ 10747~\AA \& 10798~\AA, and \fexi\ 7892~\AA\ (bottom) lines constructed from the scan observed on 2005 September 21. Plus symbols in the top two panels locate the individual loop structures chosen. Numbers given at the end of these structures, are for reference.}
\label{fig1}
\end{figure}
These images were therefore aligned and brought to same spatial scale using the wire spectra in each line. The pixel scale after this correction is found to be $\approx$ 2.2\arcsec\ per pixel for the scans in set I and $\approx$ 2.4\arcsec\ per pixel for those in set II. However, the spatial resolution in the scan direction is limited by the slit width. Final images after all these corrections, for intensities in \fexiv\ 5303~\AA, \fexiii\ 10747~\AA\ \& 10798~\AA, and \fexi\ 7892~\AA\ emission lines corresponding to the scan taken on 2005 September 21, are shown in Figure~\ref{fig1}. 

\subsection{Translating Emission Line Intensity Ratios to Temperature}
 \label{chianti_model}
Ratio of intensities from a suitable line pair can be used to get the thermal information of the source. We chose the emission line pairs (\fexiv\ 5303~\AA, \fexiii\ 10747~\AA) and (\fexi\ 7892~\AA, \fex\ 6374~\AA) to study the thermal structure of a loop, since the emission from both the lines in each pair is expected to originate from the same plasma volume due to the closeness in their maximum abundance temperatures. The temperatures of maximum abundance for the ions \fexiv, \fexiii, \fexi, and \fex\ are 2.0 MK, 1.8 MK, 1.4 MK, and 1.1 MK respectively, as noted from the atomic database CHIANTI \citep{1997A&AS..125..149D}. These line ratios are sensitive to electron temperature which means that any change in the temperature will be reflected in the ratio of the observed intensities. We modeled the dependence of these ratios on temperature using CHIANTI (version 7.1; \citet{1997A&AS..125..149D, 2012ApJ...744...99L}). The calculations are done at a constant density using the coronal abundances of \citet{2012ApJ...755...33S} and ionization fractions of \citet{1992ApJ...398..394A}. Photoexcitation from a radiation field of temperature 6000~K at a distance of 1.1~$R_\sun$ is included. A calibration curve is constructed with the modeled intensity ratio values for a range of temperatures. However, it was found that there is a significant dependence of these ratios on density as well. 
\begin{figure}
\centering
\includegraphics[scale=1.2]{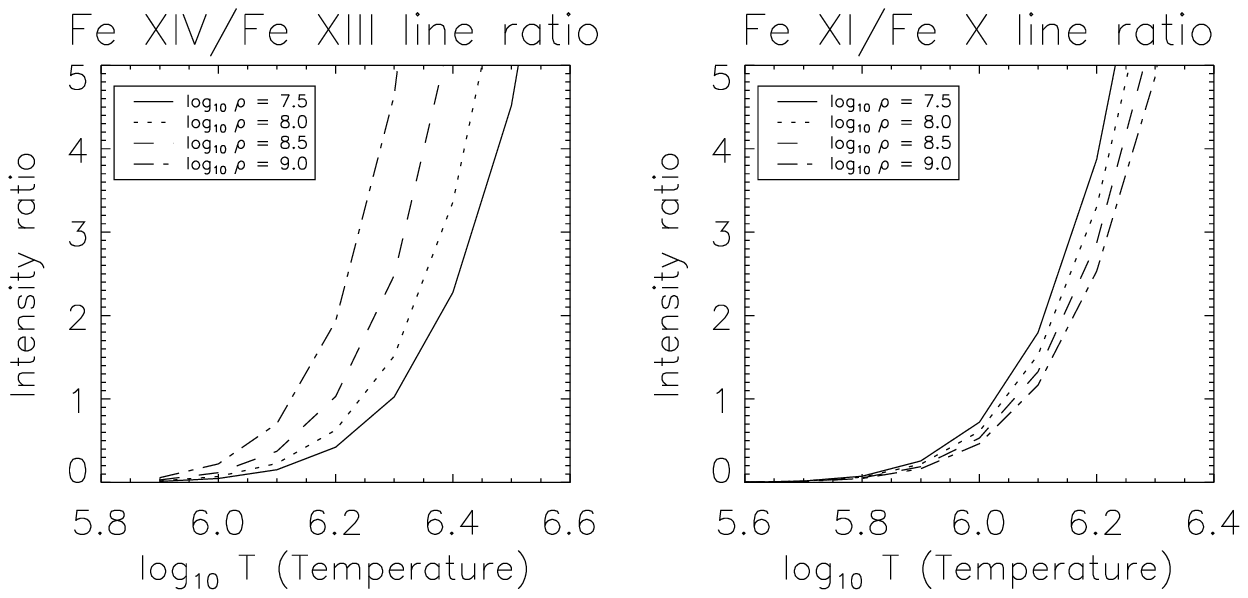}
\caption{Variation of emission line ratios \fexiv/\fexiii\ (left) and \fexi/\fex\ (right) with electron temperature modeled using CHIANTI. Different curves are plotted for different densities. See text for other parameters used in the model.}
\label{fig2}
\end{figure}
We therefore, produced similar calibration curves at different densities varying from 10$^{7.5}$ to 10$^{9.0}$ cm$^{-3}$ in steps of 0.1 in the log scale. The constructed curves for both the ratios at four density steps are shown in Figure~\ref{fig2}. The calculations by \citet{2007BASI...35..457S} indicate a weak density dependence of the \fexiv/\fexiii\ line ratio. On the contrary, our analysis using CHIANTI shows this to be significant. The density dependence of the \fexi/\fex\ ratio is relatively weaker. Figure~\ref{fig2} also shows an opposite dependence for the two ratios on density. The curves in the left panel shift towards the lower temperature side as the density increases whereas that in the right panel shift towards the higher temperature side. The temperature at any location of the scan can be estimated from these curves, using the density and the observed intensity ratio (from the appropriate line pair) at that location.

\subsection{Temperature Variation along Loop Structures}
Individual loop structures that are clearly visible in the images for both the lines of a line pair were identified. These structures were followed manually by clicking at many locations along the structure in one of the emission lines and the same pixel locations are automatically chosen from the images in other lines observed at the same time. Only those parts of the loops that are clearly discernible and seemingly isolated are chosen. Plus symbols in Figure~\ref{fig1} outline the chosen loop structures from that scan. Numbers marked at the end of these structures, are for reference. Intensity ratios were computed along each of these structures. Electron densities were derived along each loop using the observations in IR line pair \fexiii\ 10747~\AA\ \& 10798~\AA, whose ratio is sensitive to changes in density but a weak function of temperature \citep{1984SoPh...94..117N}. We verified this with CHIANTI and found the ratio to be fairly constant with $\lesssim$ 1~\% changes over a temperature range of 1 -- 4 MK. However, the signal in 10798~\AA\ line is poor and the density information could not be obtained at all the locations along a loop. So, we fit the log values of the observed densities linearly to derive the value at each pixel location along the structure. Density was found to vary between 10$^{8.5}$ to 10$^{8.0}$ cm$^{-3}$ in the chosen structures. Density information could be not obtained for the scans in set~II as they are observed only in \fexi\ and \fex. For the structures chosen from these scans, we model the density variation with the average gradient observed in the structures from set~I and taking the density at the base to be 10$^{8.5}$ cm$^{-3}$. This may have some uncertainty, but at least allows us to take the density variation into account. Combining this density information with the observed line ratios, we derived the temperature along each loop structure using the calibration curves constructed, as explained in \S~\ref{chianti_model}. These values are then plotted and fitted with a first order polynomial to obtain the gradients. We also compute the temperature values for a constant density scenario to check the effect of density variation.
\begin{figure}
\centering
\includegraphics[]{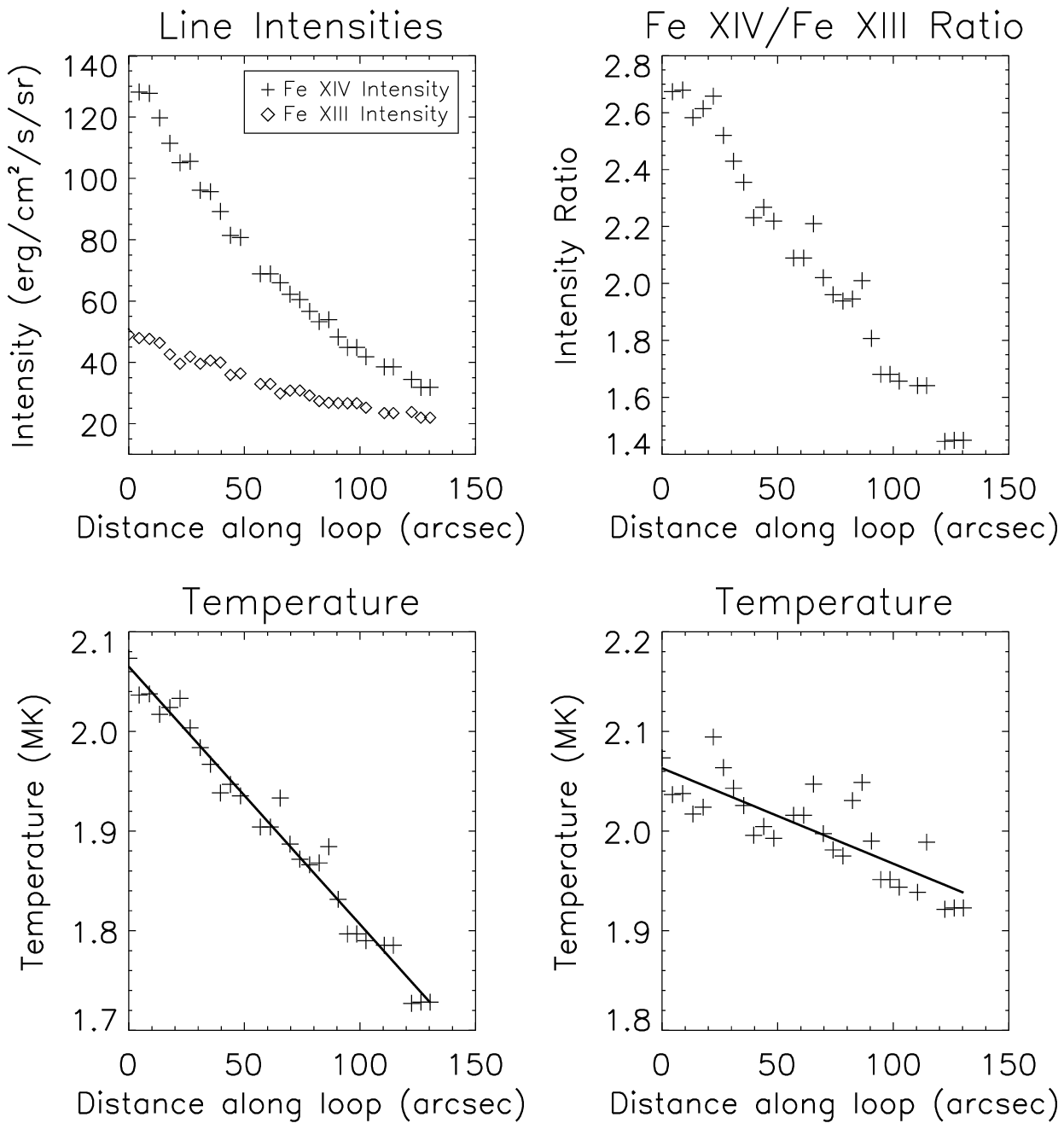}
\caption{Variation of different parameters along loop \#5 marked in Figure~\ref{fig1}. Temperature values in the bottom left plot are derived for a constant density 10$^{8.5}$ cm$^{-3}$ and that on the bottom right are derived taking the density variation into account. The solid lines over plotted on these values represent a linear fit.}
\label{fig3}
\end{figure}
Observed intensities in 5303~\AA\ and 10747~\AA\ lines, their ratio and the temperature profiles along loop \#5 (see Figure~\ref{fig1}) are shown in Figure~\ref{fig3}. Temperature values in the bottom left panel are derived for a constant density 10$^{8.5}$ cm$^{-3}$ and that on the bottom right are derived taking the observed density variation into account. The solid lines over plotted on these values represent a linear fit to the data. The scatter in temperature values increased when the density dependence of the line ratios is taken into account which could possibly arise due to uncertainties in the density estimation.
\begin{table*}
\begin{center}
\caption{Temperature values and their gradients along individual loop structures.}
\label{tbl2}
 {\small
 \begin{tabular}{clccccc}
  \tableline \tableline
  Set & Date & loop & Temperature & Temperature & $\nabla$T$_{d}$\tablenotemark{a} & $\nabla$T\tablenotemark{b}\\  
      &      & (\#) & at 10\arcsec\ (MK) & at 100\arcsec\ (MK) & (MK arcsec$^{-1}$) & (MK arcsec$^{-1}$)\\
  \tableline
   I & 2005 Sep 21 &  1 & 2.12$\pm$0.009 & 2.05$\pm$0.005 & -0.0007$\pm$0.00007 & -0.0013$\pm$0.00006 \\
     & 2005 Sep 21 &  2 & 2.23$\pm$0.006 & 2.02$\pm$0.007 & -0.0023$\pm$0.00016 & -0.0022$\pm$0.00015 \\
     & 2005 Sep 21 &  3 & 2.26$\pm$0.008 & 2.04$\pm$0.006 & -0.0024$\pm$0.00012 & -0.0033$\pm$0.00009 \\
     & 2005 Sep 21 &  4 & 2.02$\pm$0.010 & 1.92$\pm$0.010 & -0.0012$\pm$0.00018 & -0.0021$\pm$0.00020 \\
     & 2005 Sep 21 &  5 & 2.05$\pm$0.009 & 1.97$\pm$0.007 & -0.0010$\pm$0.00014 & -0.0026$\pm$0.00008 \\
     & 2005 Sep 21 &  6 & 2.05$\pm$0.008 & 2.07$\pm$0.005 &  0.0003$\pm$0.00011 & -0.0010$\pm$0.00009 \\
     & 2005 Sep 21 &  7 & 2.06$\pm$0.008 & 2.03$\pm$0.005 & -0.0003$\pm$0.00010 & -0.0011$\pm$0.00007 \\
     & 2005 Oct  3 &  8 & 2.00$\pm$0.014 & 1.93$\pm$0.009 & -0.0008$\pm$0.00018 & -0.0010$\pm$0.00014 \\
     & 2005 Oct  3 &  9 & 2.17$\pm$0.013 & 2.18$\pm$0.008 &  0.0001$\pm$0.00015 & -0.0007$\pm$0.00012 \\
     & 2005 Oct  3 & 10 & 1.93$\pm$0.012 & 1.95$\pm$0.014 &  0.0002$\pm$0.00025 & -0.0020$\pm$0.00020 \\
     & 2005 Oct  3 & 11 & 2.30$\pm$0.020 & 2.29$\pm$0.014 & -0.0001$\pm$0.00028 & -0.0007$\pm$0.00024 \\
     & 2005 Oct  6 & 12 & 2.17$\pm$0.007 & 1.95$\pm$0.009 & -0.0025$\pm$0.00019 & -0.0015$\pm$0.00017 \\
     & 2005 Oct  6 & 13 & 1.99$\pm$0.011 & 1.97$\pm$0.011 & -0.0003$\pm$0.00021 & -0.0023$\pm$0.00016 \\
     & 2005 Oct  6 & 14 & 1.99$\pm$0.010 & 1.79$\pm$0.010 & -0.0022$\pm$0.00017 & -0.0013$\pm$0.00018 \\
     & 2005 Oct  6 & 15 & 1.96$\pm$0.010 & 1.92$\pm$0.011 & -0.0005$\pm$0.00020 & -0.0014$\pm$0.00014 \\
     & 2005 Oct  6 & 16 & 1.97$\pm$0.012 & 1.89$\pm$0.013 & -0.0009$\pm$0.00022 & -0.0015$\pm$0.00016 \\
  \tableline
  II & 2007 Oct  6 & 17 & 1.21$\pm$0.011 & 1.49$\pm$0.017 & 0.0031$\pm$0.00022 & 0.0034$\pm$0.00020 \\
     & 2007 Oct  6 & 18 & 1.40$\pm$0.012 & 1.46$\pm$0.017 & 0.0006$\pm$0.00022 & 0.0008$\pm$0.00022 \\
  \tableline
 \end{tabular}}
\tablenotetext{a}{Temperature gradients taking the density dependence of the line ratio into account.}
\tablenotetext{b}{Temperature gradients for a constant density (10$^{8.5}$ cm$^{-3}$).}
\end{center}
\end{table*}
Temperature values at 10\arcsec\ \& 100\arcsec\ distances from the base of the chosen structure and their gradients are tabulated in Table~\ref{tbl2} for all the loops investigated in this study. In the last column, we list the temperature gradients in the loops keeping the density constant at 10$^{8.5}$ cm$^{-3}$, for comparison. The 1-$\sigma$ uncertainties in each of these parameters, obtained from the fit, are also given in the table.

\section{Results and Discussion}
We investigated 18 loop structures for the temperature profiles along their length using emission line ratios. Out of these 16 are from the scans in set~I and the rest two are from those in set~II. Table~\ref{tbl2} lists the computed temperature at 10\arcsec\ \& 100\arcsec\ distances along with the temperature gradient for each of these structures. The gradient values represent a significant change in temperature along the loop length (see Figure~\ref{fig3}). Also, none of the structures was observed from the foot point through the full length of a loop, so the temperature difference between the foot point and apex of the loop can be even larger. Clearly, the gradients are negative for most of the loops (positive but close to zero for the rest) under set~I and positive for those under set~II. The average gradient values are -0.0009$\pm$0.00004 MK~arcsec$^{-1}$ and 0.0018$\pm$0.00016 MK~arcsec$^{-1}$ respectively, for these two cases after accounting for the density variation along the loops. Note that the temperature estimation in the former case is from [\fexiv] 5303~\AA/[\fexiii] 10747~\AA\ ratio whereas that in the latter case from [\fexi] 7892~\AA/[\fex] 6374~\AA\ ratio. The data sample in set~II is limited (only two loops \#17 \& \#18). However, it complements the extensive analysis done by \citet{2004ApJ...617L..81S}, who finds an increase in 7892/6374 line intensity ratio along 90~\% of the structures studied, implying a positive temperature gradient. Therefore, the observed variations in temperature from [\fexiv]/[\fexiii] and [\fexi]/[\fex], possibly indicate a general behavior. The line pair (\fexiv, \fexiii) represents a slightly hotter plasma compared to the (\fexi, \fex) pair. So, it appears that the loop tops are, in general, colder when observed in hotter lines and hotter in relatively colder lines, with respect to their coronal foot points. Further, the average temperature values from both the sets are closer to each other at larger heights (100\arcsec) indicating the likelihood to reach a common value at greater heights \citep{2006SoPh..236..245S}. 

The temperature in a uniformly heated stable hydrostatic loop increases along its length and peaks at the loop top \citep{1978ApJ...220..643R}. Also, if the heating is non-uniform with a bulk of the heat deposited close to the foot points, then the temperature maximum can occur well below the loop top leading to a negative temperature gradient along its length \citep{1999ApJ...512..985A, 2001ApJ...550.1036A}. But in such cases, if the heating scale height is smaller than one-third of the loop half-length, loops become thermally unstable and are short lived \citep{1981ApJ...243..288S}. Recently, \citet{2012ApJ...755...86H} report the ubiquitous presence of such loops in the low latitude quiet corona. However, taking the generality of our results into account, it is not possible to explain the contrasting trends observed in different temperature plasma with different types of heating. Alternatively, if the loop structures studied here are composed of several unresolved strands that are impulsively heated resulting in a multi-thermal structure and if we allow the hotter and colder plasma to interact with each other gradually, the observed temperature decrease in hotter plasma and increase in colder plasma can be easily explained. Since the loops are observed here at a moderate spatial resolution ($\approx$ 2\arcsec\ per pixel), it is possible that they are multi-stranded \citep{2000ApJ...528L..45R} and multi-thermal \citep{2007ApJ...667..591P}. At this point, the exact mechanism causing the gradual interaction between the hotter and colder plasma remains unclear. But we believe that this process holds key to constrain the loop heating models and needs further investigation.

\acknowledgements 
The authors thank the referee for useful comments. J. Singh thanks the staff at Norikura Observatory for their help during observations. He also wish to thank NAOJ for providing the financial assistance for visits to make observations. CHIANTI is a collaborative project involving George Mason University, the University of Michigan (USA) and the University of Cambridge (UK).

\end{document}